\documentclass[aps,showpacs,onecolumn,floats,prd,superscriptaddress,nofootinbib]{revtex4-1} 
\usepackage{graphicx,amsmath,amssymb,amstext}
\usepackage{amssymb,amsbsy,amsfonts,amsthm,color}
\usepackage{epsfig}
\usepackage{graphicx}
\usepackage{subfigure}
\usepackage{sidecap}
\usepackage{floatrow}
\graphicspath{{Figures/}}

\begin{document}

\title{A self-consistency check for unitary propagation of Hawking quanta}

\author{Daniel Baker}
\email{dbaker@cita.utoronto.ca}
\affiliation{Canadian Institute of Theoretical Astrophysics, 60 St George St, Toronto, ON M5S 3H8, Canada.}
\affiliation{University of Toronto, Department of Physics, 60 St George St, Toronto, ON M5S 3H8, Canada.}

\author{Darsh Kodwani}
\email{dkodwani@physics.utoronto.ca}
\affiliation{Canadian Institute of Theoretical Astrophysics, 60 St George St, Toronto, ON M5S 3H8, Canada.}
\affiliation{University of Toronto, Department of Physics, 60 St George St, Toronto, ON M5S 3H8, Canada.}

\author{Ue-Li Pen}
\email{pen@cita.utoronto.ca}
\affiliation{Canadian Institute of Theoretical Astrophysics, 60 St George St, Toronto, ON M5S 3H8, Canada.}
\affiliation{Canadian Institute for Advanced Research, CIFAR program in Gravitation and Cosmology.}
\affiliation{Dunlap Institute for Astronomy \& Astrophysics, University of Toronto, AB 120-50 St. George Street, Toronto, ON M5S 3H4, Canada.}
\affiliation{Perimeter Institute of Theoretical Physics, 31 Caroline Street North, Waterloo, ON N2L 2Y5, Canada.}

\author{I-Sheng Yang}

\email{isheng.yang@gmail.com}
\affiliation{Canadian Institute of Theoretical Astrophysics, 60 St George St, Toronto, ON M5S 3H8, Canada.}
\affiliation{Perimeter Institute of Theoretical Physics, 31 Caroline Street North, Waterloo, ON N2L 2Y5, Canada.}

\begin{abstract}
The black hole information paradox presumes that quantum field theory in curved spacetime can provide unitary propagation from a near-horizon mode to an asymptotic Hawking quantum. 
Instead of invoking conjectural quantum-gravity effects to modify such an assumption, we propose a self-consistency check.
We establish an analogy to Feynman's analysis of a double-slit experiment. 
Feynman showed that unitary propagation of the interfering particles, namely ignoring the entanglement with the double-slit, becomes an arbitrarily reliable assumption when the screen upon which the interference pattern is projected is infinitely far away.
We argue for an analogous self-consistency check for quantum field theory in curved spacetime.
We apply it to the propagation of Hawking quanta and test whether ignoring the entanglement with the geometry also becomes arbitrarily reliable in the limit of a large black hole. 
We present curious results to suggest a negative answer, and we discuss how this loss of na\"ive unitarity in QFT might be related to a solution of the paradox based on the soft-hair-memory effect.
\end{abstract}

\maketitle

\section{Introduction and Summary}

\subsection{Black Hole Information Paradox}

The black hole information paradox \cite{Haw76a} was sharpened to show a self-inconsistency among the following three widely-believed statements \cite{AMPS}\footnote{Different physicists may believe in some of those more strongly than the others, but very few would have argued against any single statement without the explicit conflict in the information paradox.}:
\begin{itemize}
\item {\bf Unitary evaporation:} A black hole will totally evaporate; the formation and evaporation of a black hole is described by an asymptotic observer as a unitary $S$-matrix, whose size is the exponential of the black hole's Bekenstein entropy.
\item {\bf General relativity:} The collapse-Schwarzschild geometry given by general relativity is a valid description of the spacetime everywhere except for near the singularity.
Therefore, there is no drama while crossing the horizon.
\item {\bf Quantum field theory in curved spacetime:} Away from the singularity, we can apply local quantum field theory (QFT) which describes microscopic steps of evaporation. 
For large black holes, it manifests itself as unitary propagation of every near-horizon mode into every asymptotic Hawking quantum. The back-reaction of such an effect is no more than a Schwarzschild metric with a very slowly changing mass.
\end{itemize}
Basically, Statement Three demands that a near horizon mode is related to an asymptotic Hawking quantum by a unitary transformation, and therefore carries the same qubit of information. 
However, Statement One and Two demand this unique qubit to be purified by two distinct objects, violating monogamy \cite{Bou13}. 
This conflict has inspired many different proposals to modify one of the above three statements. 
For example, information loss or remnant \cite{Bek94} modifies Statement One; 
firewall \cite{BraPir09,AMPS}, ER=EPR \cite{MalSus13}, pre-Hawking radiation \cite{KawMat13,Ho16}, or some version of the fuzzball \cite{Mat15} modifies Statement Two;
various proposals of nonlocal effects near the horizon \cite{Gid12,DodSil15,OsuPag16}, causal patch complementarity \cite{HuiYan13,IlgYan13,LowTho14}, or computational complementarity \cite{HH} modify Statement Three.\footnote{Various versions of complementarity basically claim that QFT breaks down if one applies it to compare quantities which are in-principle not measurable (or difficult to measure) by the same observer.} 

In this paper, we will try to provide a stronger motivation to modify Statement Three. 
In fact, we will argue that it is already problematic on its own. 
In other words, {\bf the application of local QFT to unitary propagation of individual Hawking quantum is not self-consistent. } 

Before presenting our argument, we will first explain its relation to some existing ideas.
The recent paper by Osuga and Page \cite{OsuPag16} provides an abstract framework of how modifying Statement Three resolves the paradox.
Basically, the near horizon mode is purified by its interior partner as demanded by Statement Two.
However, it propagates out in a non-unitary process, losing its original state and gaining the appropriate information to restore the unitary $S$-matrix demanded by Statement One. 
For this abstract model to work in practice, this non-unitary propagation should be understood as an interaction with a hidden system, which actually exchanges the information within the out-going quantum with a qubit of information in the hidden system.
Here hidden simply means something that a na\"ive application of local QFT is ignorant to.
The obvious candidate of such hidden system is the geometry.

One proposal along this line of thoughts was discussed in \cite{NomVar12}. 
The idea was that the black hole acquires a large uncertainty during the evaporation process. 
If we treat the superposition of different classical geometries as a quantum system, it may qualify as the hidden system that is responsible for the non-unitary propagation of Hawking quanta. 
Unfortunately, the early version of this proposal can be defeated by counting entropy.
In order for such non-unitary process to restore the asymptotic $S$-matrix, the hidden system must store the required information, which is comparable to the amount of microscopic information carried by the black hole.
However, the black-hole-no-hair theorem states that two black-hole geometries with identical macroscopic parameters (mass, momentum, angular momentum, charge) are isometric to each other. 
This limits the number of different classical geometries to be parametrized by macroscopic parameters only.
As a result, the Hilbert space of superposed classical geometries has far fewer dimensions than what is necessary to carry the required amount of information.\footnote{We thank Raphael Bousso for private communication.}
Thus, even if a propagating Hawking quantum does interact with this hidden system, it cannot reproduce a unitary $S$-matrix.
Therefore such modification of Statement Three alone is insufficient to resolve the paradox.

The recent development on black-hole-soft-hairs \cite{HawPer16} hinted a way to revive this idea.
Essentially, it was realized that two isometric (regions of) geometries should not always be considered as the same state.
If one maps the geometries with congruences of geodesics, then two initially identical geometries with identical mappings may evolve into two late-time isometric geometries with different mappings. 
Since changes in the mapping geodesics are measurable memory effects, a consistent physical theory must be able to distinguish them.
Thus isometric regions with different maps should be described as different states.
This realization may dramatically increase the number of different states encoded in classical geometries.
A tentative counting of state on a black hole horizon yields exactly the required number such that a modification to Statement Three can indeed resolve the paradox \cite{HawPer16}.

In this paper, we will take another step that is complementary to the above progress. It is exciting to know that modifying Statement Three can, in-principle, resolve the paradox.
However, it remains a great mystery why local QFT should break down in low curvature regions. 
If we just demand that it breaks down in order to resolve the information paradox, then there is no reason why this solution is preferred over modifying Statement One or Two. 
We would like to argue that there is a self-consistency check that we should apply to QFT in curved spacetime, for a reason that is totally independent from the information paradox.
Applying such self-consistency check, we can show that a unitary propagation from a near horizon mode to an asymptotic Hawking quantum is already self-inconsistent. 
Thus, it is only natural to modify Statement Three.

\subsection{Quantum Field Theory in Curved Spacetime Requires a Self-Consistency Check}

The starting point of our argument involves no new physics at all.
We simply recognize that local QFT implies a standard assumption of {\bf subsystem unitarity}.
Quantum mechanical evolution of a full system is by-definition unitary. 
However in practice, we are not able to describe everything by quantum mechanics all together, so we describe only subsystems. 
Namely, we assume that the full system can be separated into a ``classical background'' and a ``quantum subsystem''. 
After this artificial separation, since the quantum subsystem is the only thing we describe by quantum mechanics, it always appears to be unitary.
However, such unitarity should not be taken as a physical fact unless one specifically checks the nature of interactions with the background.

The most direct way to justify the classical-background assumption is to also describe the background in quantum mechanics, and verify that the quantum interaction between the background and the quantum subsystem indeed does not entangle them. 
This is clearly difficult in practice. 
The very reason why we would like to apply the classical-background approximation is to avoid describing the far-too-complicated background by quantum mechanics. 
A consistency check which requires us to do so defeats the purpose. 
In particular, there are situations in which we in-principle do not know how to describe the classical background in quantum mechanics. 
Our problem at hand, QFT in curved spacetime, is exactly this annoying situation. 
Since the classical background is the geometry, one in-principle needs to know quantum gravity in order to model the quantum interaction with the background. 

Fortunately, Feynman, in his famous lectures, presented an interesting trick to circumvent this obstacle. 
This trick, to a certain extent, enables us to perform a {\bf self-consistency check} of the classical-background assumption {\bf without} knowing the quantum nature of the background.
In Sec.\ref{sec-DoubleSlit}, we will review Feynman's analysis of the double-slit experiment. 
In his case, the double-slit is the classical background, the particle passing through it is the quantum subsystem, and subsystem unitarity is checked by whether an interference pattern can be observed on the final projection screen. 
He showed that we can ignore the quantum details of the double-slit and summarize that as an uncertainty in its position, which is a classical quantity.
\footnote{Here we mean that an uncertainty in position is a quantity that one can directly plug into the framework/equations of a classical theory, as oppose to states or operators which can only be understood by a quantum theory.} 
When such uncertainty is fixed, a classical calculation can show that the interference pattern is always visible as long as we put the final projection screen to be infinitely far away. 
It is in this sense that a unitary evolution of the subsystem is an arbitrarily good approximation.

The fundamental principle behind Feynman's trick is that ``classical wave coherence'',
\footnote{We remind the readers that ``coherence'' has a classical (likely original) meaning as superpositions of waves with stably similar phases.}
 in his case the observability of the interference pattern, should be taken as the prerequisite of ``quantum unitarity''. 
In Sec.\ref{sec-QFT}, we argue that this principle enables a general self-consistency check of the classical background assumption. 
We provide a natural generalization to local QFT in curved spacetime, and we discuss why such self-consistency check is different from perturbative calculations of graviton-scattering.

In Sec.\ref{sec-BlackHole}, we apply the self-consistency check to the propagation of Hawking quanta in a close analogy to Feynman's analysis. 
The classical geometry plays the role of the double-slit, a background that can potentially entangle with the Hawking quanta wavefunction and ruin its subsystem unitarity. 
We then assume that the unknown quantum nature of geometry can be parametrize by a classical uncertainty with an unknown but fixed, gauge-invariant size. 
A classical wave plays the role of the interference pattern. 
When even a classical wave decoheres due to the uncertainty of geometry, there is no reason to believed that the quantized version of such wave has unitarity.
The common expectation is that in the limit of a large black hole, gravitational effects outside the horizon are arbitrarily weak, thus local QFT should be arbitrarily trustworthy.
By analogy to Feynman's analysis, we should expect that with a fixed uncertainty in the geometry, any effect on the classical wave should drop to zero when we take the large black hole limit.
Interestingly, what we found seems to be the opposite. 
We show that with a fixed geometric uncertainty, a classical wave suffers an {\it arbitrarily large} correction in the large black hole limit.
By analogy to Feynman's analysis, this is strong evidence that the apparent unitarity of QFT is, in this case, not a valid assumption.

In Sec.\ref{sec-dis}, we discuss various implications of our finding. 
First of all, our self-consistency check should be generally applicable to any geometry, and it does not invalidate local QFT entirely. 
Even when unitarity is lost in local QFT, the values of many observables do not have to change. 
Most existing applications of QFT in curved spacetime actually only care about the expectation value of particle numbers, which can be unaffected by the loss of subsystem unitarity. 
Secondly, our finding resolves the information paradox by selecting one culprit among the three statements, but it does not yet resolve the information transfer puzzle. 
We argue that our finding naturally implies an information exchange between a QFT quantum and the local geometry \cite{OsuPag16}.
The reason why local geometry carries the appropriate information to give in the first place may be related to the soft-hair proposal \cite{HawPer16}. 
Exactly how such exchange of information happens is a topic for future works.

\section{Feyman's analysis on double-slit experiments}
\label{sec-DoubleSlit}

Double-slit interference is the signature experiment to demonstrate the nature of quantum mechanics. Figure 1 shows a simple example of such experiment.
When a group of particles of the same momentum pass through slit 1 only, they reach the final screen as some probability distribution $P_1(y)$. 
When they pass through slit 2 only, they reach the final screen as a probability distribution $P_2(y)$.
When both slits are open, the probability distribution we find on the final screen is not a simple sum; $P_{12}(y) \neq P_1(y) + P_2(y)$.
Instead, the final screen shows an interference pattern which can only be explained by wavefunctions.
$P_i = \langle \phi_i | \phi_i \rangle$ and $P_{12} = (\langle \phi_1|+\langle \phi_2| ) (|\phi_1\rangle + | \phi_2 \rangle)=P_1 + P_2 + 2{\rm Re}\langle\phi_1|\phi_2\rangle$.

\begin{figure}[h!]
\begin{center}.
\includegraphics[scale = 0.5]{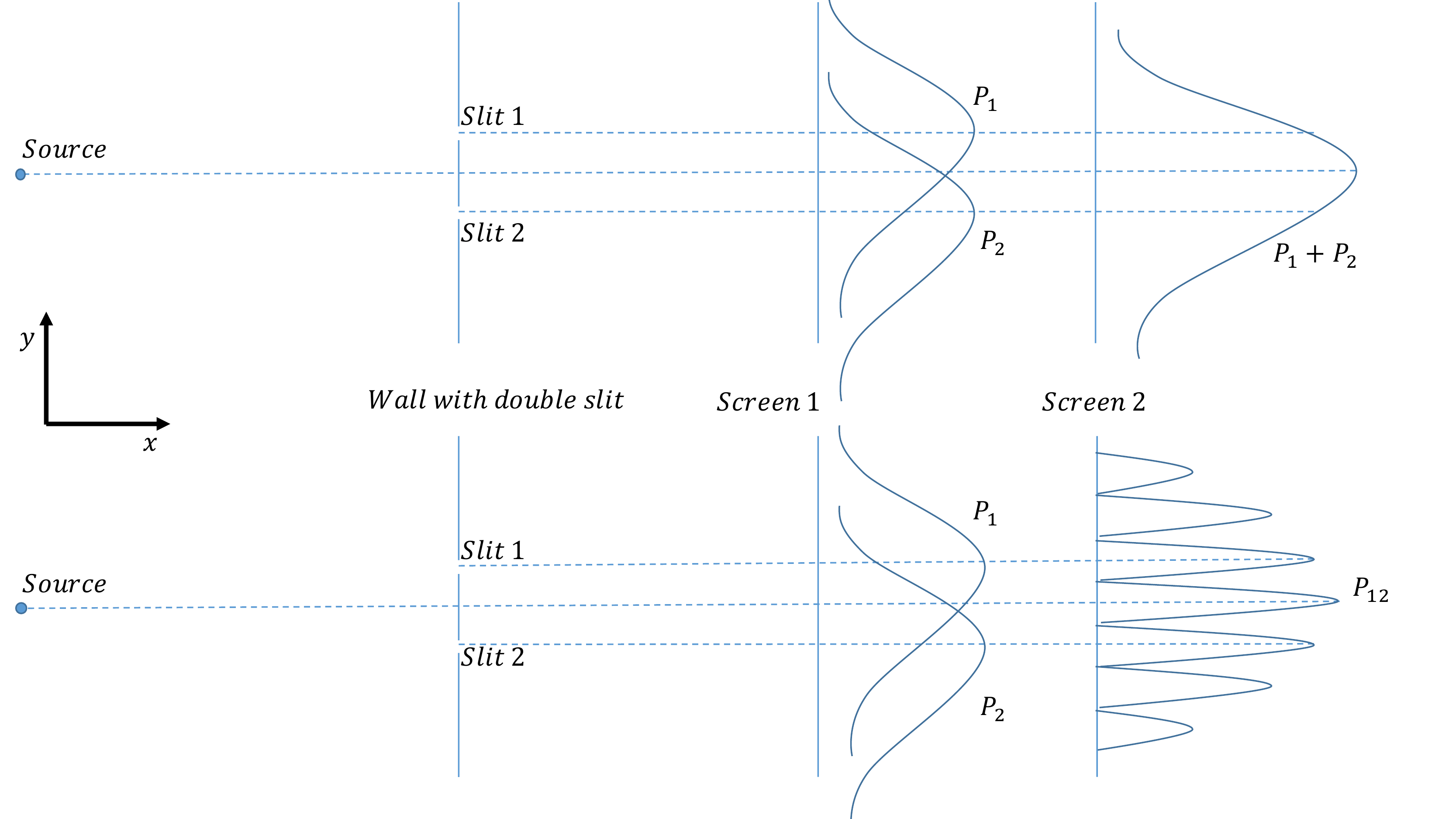}
\caption{Schematic of the double slit experiment. 
The diagram shows two different scenarios. 
The top figure shows the behaviour one would expect of classical particles going through one slit or the other. 
The bottom figure shows the behaviour of classical waves creating an interference pattern. 
Screen one shows the signal that one would see in each scenario (i.e classical particles or waves) if only one slit was open. 
If slit 1 was open we would see $P_1$ and so on. Screen 2 shows the signal we see if both the silts are open in each scenario.}
\label{fig-doubleslit}
\end{center}
\end{figure}

This standard explanation assumes the particle is the only quantum-mechanical system here, and it can be describe by a pure state, $|\phi\rangle_{\rm particle} = |\phi_1\rangle + |\phi_2\rangle$.
But if quantum mechanics is correct, this particle is not the only thing to be described by a wavefunction. 
For example, we can also describe the double-slit by its wavefunction $|\psi\rangle_{\rm ds}$. The total wavefunction of the combined system can be in a pure state,
\begin{equation}
|\Psi\rangle_{\rm combined} = |\psi_1\rangle_{\rm ds}|\phi_1\rangle_{\rm particle} 
+ |\psi_2\rangle_{\rm ds}|\phi_2\rangle_{\rm particle} ~,
\end{equation}
but either subsystem does not have to be pure.

When the two double-slit states are almost indistinguishable,
\begin{equation}
\langle\psi_1|\psi_2\rangle\langle\psi_2|\psi_1\rangle \approx 1~,
\label{eq-pure}
\end{equation}
then the combined system factorizes,
\begin{equation}
|\Psi\rangle_{\rm combined} \approx |\psi_1\rangle_{\rm ds}
\left(|\phi_1\rangle_{\rm particle} + |\phi_2\rangle_{\rm particle}\right) ~.
\end{equation}
The particle subsystem indeed stays as a pure state and there will be an interference pattern.

On the other hand, if the two double-slit states are distinguishable,
\begin{equation}
\langle\psi_1|\psi_2\rangle\langle\psi_2|\psi_1\rangle \ll 1~,
\label{eq-mixed}
\end{equation}
that means the interaction entangled the two systems. The particle subsystem becomes a mixed state,
\begin{equation}
\rho_{\rm particle} = {\rm Tr}_{\rm ds}|\Psi\rangle\langle\Psi| \approx
|\phi_1\rangle\langle\phi_1| + |\phi_2\rangle\langle\phi_2|~,
\end{equation}
and there will be no interference pattern.

Whether the double-slit states are given by Eq.~(\ref{eq-pure}) or (\ref{eq-mixed}) seems to require the knowledge about the actual quantum-mechanical interaction between the particles and the double-slit, which is unavailable in practice.
Feynman pushed the above analysis further to overcome such a problem.
He pointed out that even if we try to keep the interaction minimal, there will be one inevitable interaction that makes $|\psi_1\rangle$ and $|\psi_2\rangle$ different.
That is because when a particle reaches some place on the screen, its $y$-momentum must be different depending on which slit it passed through.
\begin{equation}
\Delta p_y^{\rm particle} \equiv 
|\langle\phi_1|p_y|\phi_1\rangle - \langle\phi_2|p_y|\phi_2\rangle| 
\approx p_x \frac{s}{L}~,
\end{equation}
where $p_x$ is the $x$ momentum of the particles, $s$ is the separation between the two slits, and $L$ is the distance to the screen. 
This means two different values of recoil momentum on the double-slit,
\begin{equation}
\Delta p_y^{\rm ds} = \Delta p_y^{\rm particle} \approx p_x \frac{s}{L}~.
\label{eq-dpds}
\end{equation}
The uncertainty principle sets a limit on how well we can measure this difference. 
Namely, we need a large uncertainty in position to measure a small change in momentum. 
Setting $\hbar$ to 1, we can say that when the momentum difference is too small to be measured,
\begin{equation}
\Delta p_y^{\rm ds} < 1/\Delta y^{\rm ds}~,
\label{eq-standard}
\end{equation} 
then Eq.~(\ref{eq-pure}) is true and the two states are indistinguishable. 
Otherwise Eq.~(\ref{eq-mixed}) is true and the two states are distinguishable.

The key point is that there is an alternative way to appreciate Eq.~(\ref{eq-standard}) {\it without} thinking about the quantum mechanics of the double-slit. 
First of all, the interference pattern has a fringe width of  
\begin{equation}
w = L\frac{\lambda}{s}~,
\end{equation}
where $\lambda = p_x^{-1}$ is the de Broglie wavelength of the particle. 
This is the separation of bright/dark lines in the $y$ direction with a fixed reference point at the $y$ position of the double-slit. 
Thus the uncertainty in the $y$ position must be smaller than this value, so the interference pattern is not totally blurred.
\begin{equation}
\Delta y^{\rm ds} < w = \frac{L}{p_xs}~.
\label{eq-stdcls}
\end{equation}
This is exactly the same condition as in Eq.~(\ref{eq-standard}).

We should emphasize the key value of this result. 
We can treat both the position uncertainty $\Delta y^{\rm ds}$ and the interference fringe width $w$ as classical quantities. Eq.~(\ref{eq-stdcls}) is then a purely classical calculation to compare these two quantities, which tells us whether a classical wave pattern (the interference pattern) loses coherence. 
If we only take its face value, it seems to be only a practical obstacle of measuring the interference pattern. 
One might argue that the underlying subsystem unitarity is still valid, just difficult to measure. 

What Feynman showed by this example is that Eq.~(\ref{eq-stdcls}) tells us exactly the same thing as Eq.~(\ref{eq-standard}). 
The later {\bf is} directly about quantum mechanics and shows how the subsystem unitarity can fail. 
Thus an apparently classical calculation can tell us something about the hidden quantum-mechanical nature of the interaction between the double-slit and the particle passing through it. 
In this example, it tells us whether the particles get entangled with the double-slit and loses its subsystem unitarity. 

One might object that the Eq.~(\ref{eq-stdcls}) is not entirely classical, since the value of $\Delta y^{\rm ds}$ has to originate from the unknown quantum mechanics of the double-slit. 
This is a valid concern, and it actually shows another strength of Eq.~(\ref{eq-stdcls}). 
Indeed we do not know the value of $\Delta y^{\rm ds}$ just from classical physics, but it is very reasonable to assume that it is intrinsic to the double-slit. 
Namely, its value should be fixed if we move the screen further away. Eq.~(\ref{eq-stdcls}) shows that with a fixed $\Delta y^{\rm ds}$, we can always make $L$ large enough to satisfy this condition. 
Thus, at least at the level of gedanken experiment, one can always arrange a situation that the interference pattern is visible and subsystem unitarity is valid.

\section{Analogy in Quantum Field Theory}
\label{sec-QFT}

The double-slit experiment is an example where both the classical coherence and quantum unitarity can be calculated explicitly and shown to follow the same condition.
Here, we advocate that such a relation is actually a general rule.
{\it Classical coherence, instead of thinking of it as a practicality of measurements, should be treated as a prerequisit to quantum unitarity.} 
Thus when classical coherence fails, not only have we no practical measurements to confirm the purity of the subsystem wavefunction, such wavefunctions {\it actually did not evolve unitarily.}
Whatever effect responsible for a classical uncertainty large enough to disrupt classical coherence must have also interacted with the subsystem quantum-mechanically and became entangled with it.

For quantum field theory in a curved spacetime, we always assume that the quantum field is a subsystem that never entangles with the geometry, and so remains unitary.
Since we do not know quantum gravity/geometry, we cannot directly check this assumption by an actual calculation of entanglement.
However, we can always check classical coherence and, by this rule we advocate, the answer directly proves/disproves subsystem unitarity.

\begin{figure}[tb]
\begin{center}.
\includegraphics[scale = 0.5]{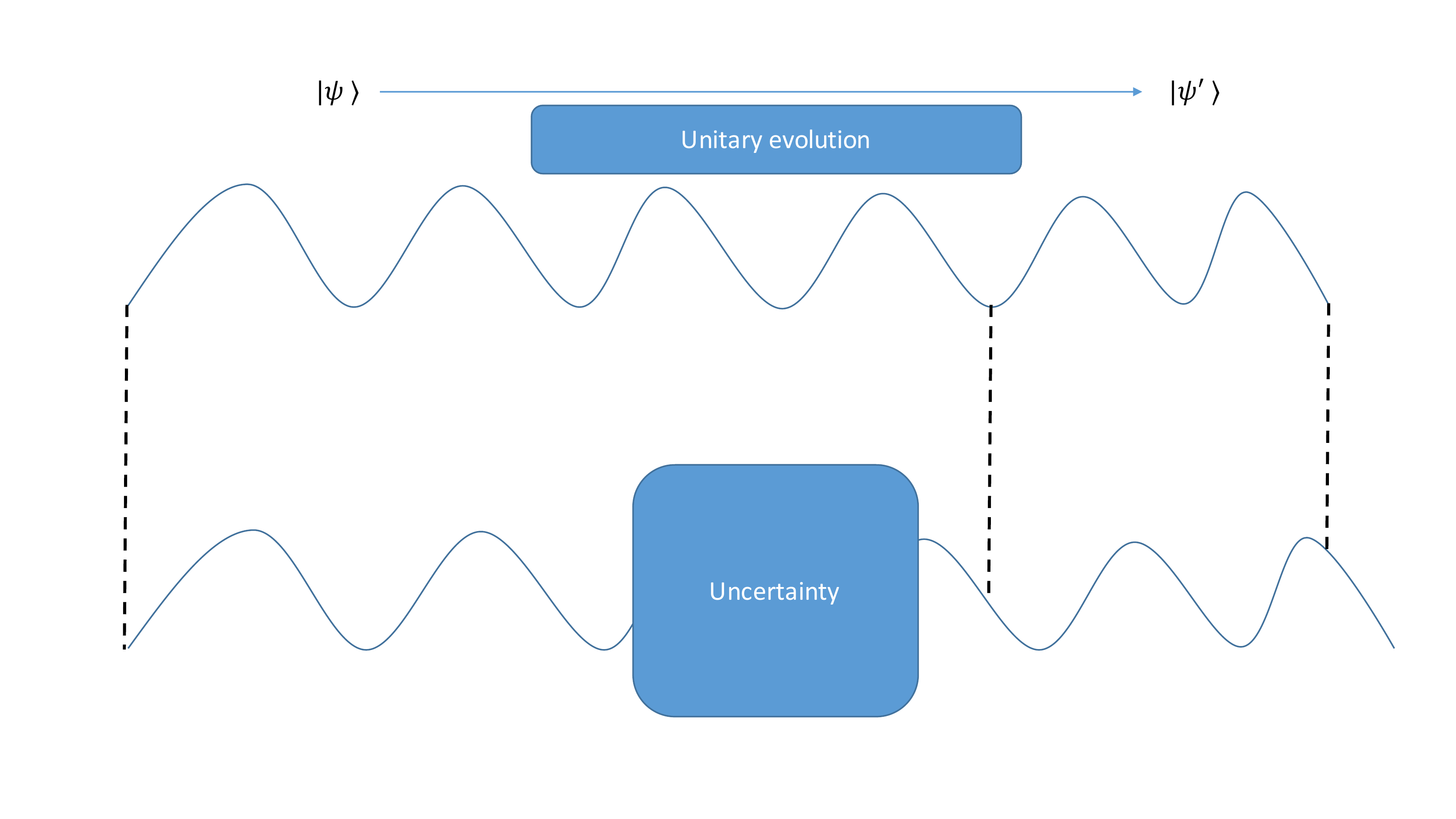}
\caption{Schematic showing the self-consistency check we advocate. 
If there is a large enough uncertainty in the classical geometry, a classical mode function can have a significant uncertainty in its phases.
If this happens, classical waves lose coherence, and the evolution of their quantum states lose unitarity.}
\label{fig-QFT}
\end{center}
\end{figure}

If we go back to the basics of quantum field theory, there is a natural description of this new rule. 
As shown in Fig.\ref{fig-QFT}, QFT starts from solving a classical mode function, then quantizing its amplitude into a quantum state. 
Any method to measure such quantum state assumes the knowledge of the classical mode function. 
If the geometric uncertainty leads to a significant uncertainty in the classical mode function, then there is no practical way to reliably measure its quantum state. 
We advocate that this is not only a practicality about measuring the quantum state, but it directly tells us that the quantum state loses its unitarity. 
Just like in Feynman's example where the uncertainty of the double-slit guarantees its entanglement with the particles, whatever effect that leads to the geometric uncertainty here must entangle with the quantum state of this mode. 
Without a theory of quantum geometry, we cannot describe how that happens. 
But through a classical calculation, we can determine whether it has happened or not.

While such generalization to QFT seems straightforward in theory, there are a few subtleties in practice.
In the double-slit experiment, the position uncertainty of the double-slit was the only obvious uncertainty of the background that we need to keep track of.
It is also the reasonable thing to hold fixed while moving the projection screen away.
In QFT, the background is the entire geometry, and there are infinitely many ways a geometry can be uncertain.
For a full analogy to Feynman's analysis, not only do we need to parametrize those uncertainties, we also need to choose the appropriate combination to hold fixed as we take a similar limit.
At this moment, we do not have a general formalism to setup such self-consistency check that can be applied to any general geometry. 
In Sec.\ref{sec-BlackHole}, we propose a specific way to apply such a self-consistency check to the propagation of Hawking quanta.
In addition to gaining insights into the information paradox, we hope to learn some lessons to inspire a more general formalism of such self-consistency checks.

QFT theorists might think that the more direct self-consistency check is feasible.
Instead of parametrizing a classical geometric uncertainty, one can treat deviations from the background geometry as a field and quantize it as any other fields.
Then a standard QFT calculation of scattering with this new field (graviton) should capture all interactions with the geometry, and a small scattering amplitude should guarantee that the interaction is suppressed. 
The concern for this method is that IR issues for these scattering calculations cannot be unambiguously regulated as in global Minkowski space.
Since soft gravitons lead to real physical changes in the geometry that can be seen in memory effects \cite{Wei65,HeLys14}, an IR ambiguity signals our ignorance on how these geometric changes enter QFT.
Thus a direct graviton scattering calculation cannot be the full extent of a self-consistency check, and it is reasonable for our method to give extra constraints on the reliability of local QFT.

\section{Hawking quanta propagation}
\label{sec-BlackHole}

The usual treatment of how a near-horizon mode becomes an asymptotic Hawking quantum also uses a fixed-background assumption.
The background is the Schwarzschild geometry.
\begin{equation}
ds^2 = -\left(1-\frac{2M}{r}\right)dt^2 + \frac{dr^2}{1-\frac{2M}{r}} + r^2d\Omega_2^2~.
\label{eq-metric}
\end{equation}
When one applies local quantum field theory to describe a field on this fixed background, the results always appear to be unitary, but that is not necessary a physical fact. 
Given how little we understand quantum gravity, a classical self-consistency check as we described in the previous section is a reasonable thing to do.

\subsection{Setup and summary}

\begin{itemize}
\item {\bf Classical coherence by tracking geodesics.} The classical mode of a massless field has its peaks and nodes following null geodesics.
Thus, instead of actually solving the mode function, we can perform a much simpler calculation to track geodesics.
We start with two out-going null geodesics with the separation of one period of such mode near the horizon, and then we calculate the change in their separations at spatial infinity.
As this change increases, one should gradually lose faith in the unitary propagation of such a mode, since it becomes increasingly difficult to experimentally verify.
\item {\bf The background and the uncertainty.} The classical background in this case is the Schwarzschild geometry, which can be written as the metric $g_{\mu\nu}$ in Eq.~(\ref{eq-metric}). 
A classical uncertainty can be expressed as deviation from such metric, $\Delta g_{\mu\nu}$. 
There are two major challenges here:
\begin{enumerate}
\item There are many more variables than the double-slit experiment. 
We consider only one particular form of $\Delta g_{\mu\nu}$ in this paper---back-reactions from the presence of extra matter of zero total energy. 
It is simple to calculate and reflects the physical intuition of vacuum pair fluctuation.
\item The value of $\Delta g_{\mu\nu}$ is gauge dependent. 
By relating it to the presence of extra matter, we can express it as a gauge-invariant local quantity, which is the natural thing to hold fixed in our analysis.
\end{enumerate}
Note that if we were reporting a positive result, that such uncertainty upholds unitarity, then one should question the validity of learning a general lesson from a special example. However, we will show that this particular form of $\Delta g_{\mu\nu}$ is already sufficient to question unitarity, and we did not choose this form to specifically do that. Thus our limited analysis is sufficient to raise a reasonable doubt to the classical-background approximation.
\item {\bf Renormalization.} It is well-known that na\"ive applications of QFT often suffers from UV/IR divergences.
In our case, we will see that even in Minkowski space, geometric uncertainty already leads to a finite change between the two null geodesics.
We will simply ``renormalize'' that away by saying that if the black hole case leads to a similar change, it should not be taken as a serious problem.
Our surprising result is that in the black hole case, we get an infinitely larger change, which is difficult to blame on the usual divergence of QFT.
\end{itemize}

\begin{figure}[tb]
\begin{center}
\includegraphics[scale = 0.6]{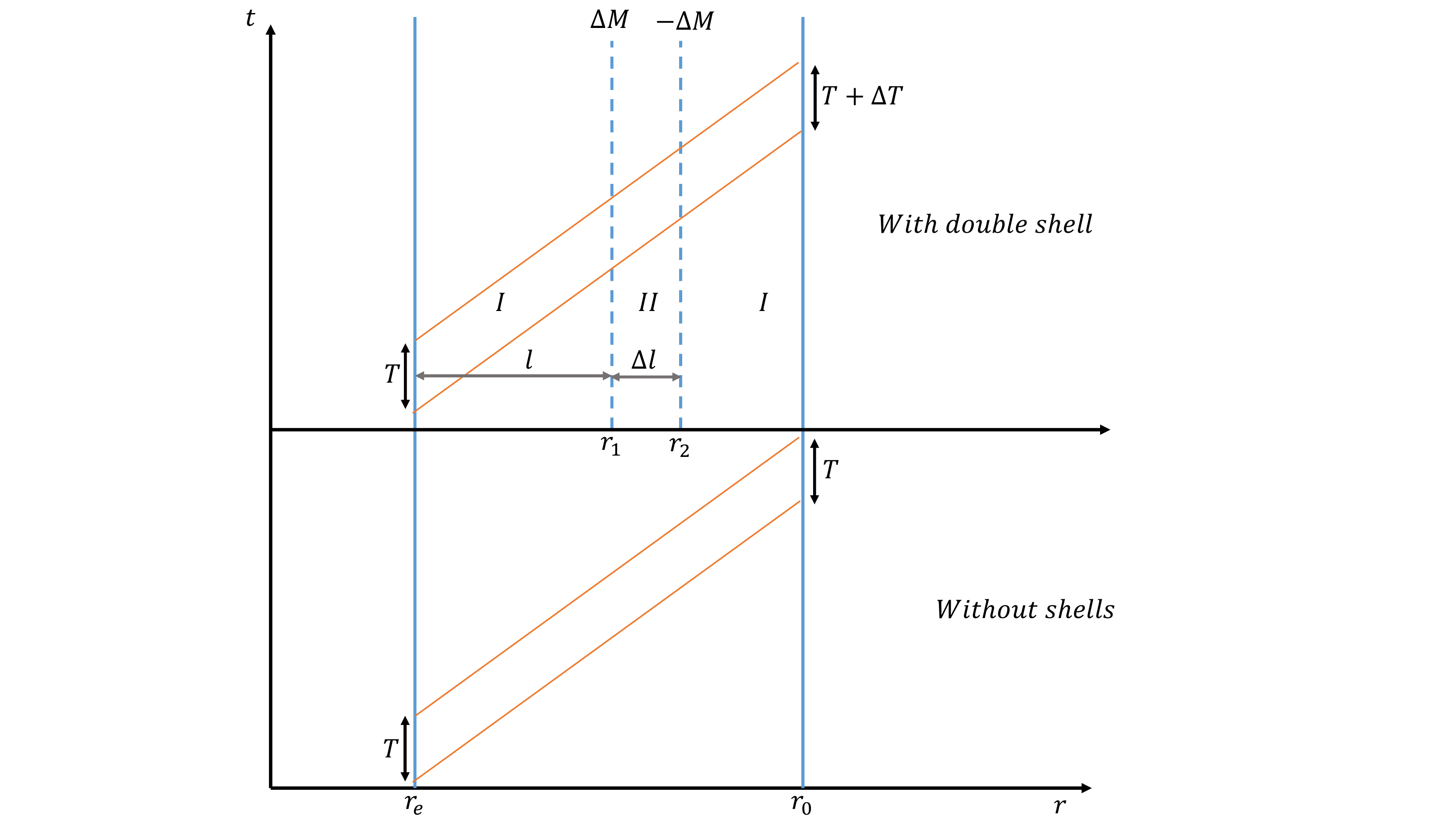}
\caption{The two parts of this diagrams describe two different scenarios. The trajectories of the photon are the lines in orange. The trajectory of the observer and the position from which the photons are emitted are represented by solid blue lines. The bottom part is a general case of two Hawking quanta coming from a distance $r_e$ from the black hole of mass $M$ and arriving at an observer who is at a distance of $r_0$. The top part shows two Hawking photons coming from the same place very close to the horizon, $r_e\approx2M$, from a black hole of mass $M$. Instead of the photons freely propagating through to an observer at $r_0$, they have to cross two shells, represented by blue dotted line, of equal and opposite mass $\Delta M, -\Delta M$ at $r_{1}, r_{2}$ respectively. The two regions marked by $I$ are described a metric with mass $M$. The region $II$ between the two shells is described by a metric with mass $M+\Delta M$. The proper distance from the shells to the horizon is $l$, and the proper distance between the shells is given by $\Delta l$. }
\label{fig-setup}
\end{center}
\end{figure}

More concretely, our setup is shown in Fig.\ref{fig-setup}. 
We start by choosing two out-going null rays in the background Schwarzschild spacetime given by Eq.~(\ref{eq-metric}).
Far away from the black hole, at $r_0\gg M$, they are separated by time $T \sim M$, which is one period of an asymptotic Hawking quantum.
Near the horizon, at $r_e$ with $(r_e-2M)\ll 2M$, the same two null rays will have the same coordinate time separation, but a much smaller proper time separation.
This represents a near-horizon mode.

Next, we introduce geometric uncertainty as sourced by two shells of zero total ADM mass.
Both the near horizon and the asymptotic regions are not affected by such uncertainty.
Thus it is a well-defined question to ask how the same near-horizon mode will end up when they reached the same large $r_0$. 
It turns out that the time separation between these two null rays will be different. 
Following the calculation in Appendix \ref{Dtime}, we arrive at this simple expression:
\begin{equation}
\frac{\Delta T}{T} = 
\left(\sqrt{\frac{r_2-2M}{r_1-2M}}\sqrt{\frac{r_1-2(M+\Delta M)}{r_2-2(M+\Delta M)}}-1\right)~,
\label{eq-result}
\end{equation}
where $(T+\Delta T)$ is the new separation with the geometric uncertainty, $r_1$ is the location of the inner shell, $r_2$ is the location of the outer shell, and $\pm\Delta M$ are their individual ADM masses. 

Note that we have assumed two stationary and spherically symmetric shells to simplify the calculation, but that is not the actual physical picture.
There is no reason why vacuum fluctuation suddenly manifests two complete shells together to surround the black hole, and holding them at a fixed location will require unphysically large pressure if they are close to the horizon.
As it will become clear later, our result only cares about the local energy density of two small pieces of shells at the ``location and time'' where the pair of null rays actually pass through.
It does not care about whether the whole shell exists or not, and the pressure does not directly affect the answer.
Our result is also invariant under radial boosts, thus it does not matter how the shells move before and after the crossing, or whether the shells are actually stationary.

\subsection{Physical interpretation}

In the rest of this paper, we will stay in the regime that $r_1, 2M, (r_1-2M) \gg \Delta M, \Delta r$, where $\Delta r \equiv(r_2-r_1)$. 
This guarantees that the geometric fluctuation is small, as it only affects a small region $\Delta r$, and the change in the metric $\Delta g_{\mu\nu}/g_{\mu\nu}$ is small in the affected region.
Under this assumption, we can expand Eq.~(\ref{eq-result}) to get 
\begin{equation}
	\frac{\Delta T}{T} = \frac{\Delta M \Delta r}{(r_1-2M)^2} = \frac{\Delta M \Delta r}{r_1^2} g_{tt}^{-2}~.
		\label{lim-result}
\end{equation}
Here $g_{tt}=(1-2M/r_1)$ is the time component of the metric in the background geometry, evaluated at the location where the null rays pass through the geometric uncertainty. 

It is more enlightening to write this expression in terms of local physical quantities (see Appendix \ref{finalEq} for derivations).  Let $\sigma$ be the energy density of the shell in its rest frame, and $\Delta l$ be the physical distance between the shells in that frame, we get
\begin{equation}
	\frac{\Delta T}{T} = 4 \pi \sigma \Delta l g_{tt}^{-1}~.
	\label{eq-inv}
\end{equation}
Here we can see that the effect is due to the local energy density which the null rays pass through.
Also, the combination $(\sigma\Delta l)$ is invariant under radial boosts, thus Eq.~(\ref{eq-inv}) does not care about whether the shells are actually at rest, nor in which frame we calculate it. \footnote{When the shells are moving, we will reach the same gauge-invariant answer by the following process.
First we pick an (arbitrary) frame.
$\Delta l$ is the proper distance between the two shells at the time of the first crossing in this frame.
$\sigma$ is the energy density measured in this frame.
The combination $(\sigma\Delta l)$ is again invariant.
Interesting, if the two shells are not moving in the same velocity, the result will depend on their relative velocity, and we cannot appreciate the physical meaning yet.
A more general framework to quantify the geometric uncertainty is a goal in future works.
}

The physical meaning of Eq.~(\ref{eq-inv}) is quite obvious. 
Geometric uncertainty creates a fixed-size uncertainty in the local proper time where it occurs.
If it happens in a region where the wavelength of a mode is blue-shifted, then it gets magnified to be relatively larger.
To be more precise, we evaluate Eq.~(\ref{eq-inv}) in three regimes; 
1) The Minkowski limit: $M=0$; 
2) Small black holes / far from horizon: $M \ll r_1$; 
3) Large black holes / near horizon: $r_1\gg M$. 
Instead of using the coordinate $r_1$, we will express the answer with the physical distance $l$ between the horizon and the shells. 
The relation between $r$ and $l$ can also be found in Appendix \ref{finalEq}. 
We will hold $l$, $\Delta l$ and $\sigma$ fixed while the black hole mass is being varied.

\begin{equation}
\frac{\Delta T}{T} ~ \begin{cases}
= 4 \pi \sigma \Delta l~	 &\ \ \ \ \ (M = 0)\\
\approx 4\pi \sigma \Delta l \left( 1 + \frac{2M}{l} \right) & \ \ \ \ \ (M \ll l)	\\
\approx 4 \pi  \sigma \Delta l \frac{16M^2}{l^2} & \ \ \ \ \ (M \gg l).
\end{cases}	\label{fin-result}
\end{equation}

\begin{itemize}

\item \textbf{Minkowski limit ($M=0$)}: 
The na\"ive interpretation of the $M=0$ result is that even in Minkowski space, geometric uncertainty can potentially decohere classical mode functions.
By our definition, that poses some threat to QFT in Minkowski space.
We are going to assume that unitarity in QFT is fine in Minkowski space, effectively ``renormalize away'' the effect at $M=0$.
One can imagine a simple subtraction by a counter term.
Or alternatively, since the actual value of $(\sigma \Delta l)$ is unknown, we assume that it is small enough to not cause any concern.

\item \textbf{Far from horizon ($M \ll l$)}: 
If we send signals from the Earths' surface to somewhere far in the space, this result tells us how much we can trust a unitary QFT description.
It starts to show a small deviation from the Minkowski result, but in the $M \ll l$ limit, it is almost indistinguishable. 

\item \textbf{Near horizon ($M \gg l$)}: 
If we consider a region of size $l$ across the horizon of a large black hole $M\gg l$, QFT believes that such region is locally the same as empty Minkowski.
As long as $l$ is much larger than the UV cut-off scale of the QFT, one would believe that modes in this region propagates out to become Hawking quanta, and QFT guarantees the unitarity of such process.
Here we hold $(\sigma\Delta l)$ fixed at the same value of empty Minkowski space, we can see that the effect of geometric uncertainty gets arbitrarily larger than in the Minkowski space.
Note that the absolute value of the effect cannot become arbitrarily large, since it is actually bounded at order one by the validity constraint of our calculation: $\Delta g_{\mu\nu} \ll g_{\mu\nu}$.
However, seeing that it can be arbitrarily larger than the Minkowski result should be a sufficient reason to raise a reasonable doubt about the unitarity of the propagation process.

\end{itemize}

\section{Discussion}
\label{sec-dis}

\subsection{The self-consistency check of QFT}

In this paper, we proposed a self-consistency check of QFT in curved spacetime geometries.
We have shown that propagation from the near-horizon region of a large black hole seems to fail such a check.
As explained in Sec.\ref{sec-BlackHole}, checking for consistency requires us to parametrize the uncertainty in the classical geometry.
We have chosen a particular form of uncertainty in our explicit calculation.
Our result is gauge invariant and makes physical sense, but there might be a more general way to parametrize such an uncertainty.

In hindsight, the physical meaning of our effect, Eq.~(\ref{eq-inv}), is quite clear.
The geometric uncertainty, parametrized by $(\sigma\Delta l)$, leads to a fixed physical change locally.
When there is a large gravitational redshift, the effect is amplified.
In the large black hole limit, starting from a fixed physical distance from the horizon, the redshift gets arbitrarily large, so the effect gets amplified by an arbitrarily large factor.
It is natural to expect similar effects in other geometries with horizons or even just expansions.
We look forward to exploring those more general cases.

We should emphasize that our check does not invalidate QFT all together.
We simply suggest that the QFT degrees of freedom lose their subsystem unitarity.
In the double-slit experiment, with or without the interference pattern, the total number of particles that passes through the double-slit is not affected, and the expectation value of the particle location remains in the center.
The point is that we can always find a pair of pure and mixed density matrices which give identical expectation values to many observables.
It is not unexpected that the only affected observable is the one explicitly checking unitarity.
Thus we can still rely on QFT to calculate the values of other observables.
For example, the particle number and energy spectrum of Hawking radiation can remain the same as the conventional result.

Many existing results of QFT in curved/dynamical geometries are about particle numbers, such as the particle-production calculation following a Bogoliubov transformation. 
As an example, one can easily see that
\begin{equation}
\rho_{\rm pure} = \left(\sum_n a_n|n\rangle\right)
\left(\sum_n \langle n| a_n^*\right)
\end{equation}
and
\begin{equation}
\rho_{\rm mixed} = \sum_n |a_n|^2 |n\rangle\langle n|
\end{equation}
give the same particle number. 
Thus there is no general crisis if we admit that QFT sometimes loses subsystem unitarity. 

As far as we know, an observable that explicitly checks the subsystem unitarity of QFT during particle-production events is the cosmological Bell inequality \cite{Mal15}. 
One might be able to apply a similar self-consistency check to such a model and update the prediction of whether a violation of Bell inequality is actually expected.

\subsection{The information paradox}

While formulating the information paradox, it is customary to take the large black hole limit.
Technically, it means that we first select a fixed length scale $l$ that is much longer than the UV-cutoff scale of the theory, and study a near-horizon region of this size.
Such a region will be almost identical to a region of the same size in global Minkowski space, since all locally measurable curvature effects are suppressed by $(l/M)$.
If we make the black hole arbitrarily large, $(l/M)$ goes to zero, then for all the field theory modes within this region, we are confident that they start in local vacuum states of QFT as if they live in Minkowski space.
This large black hole limit is the key argument to eliminate many na\"ive solutions to the information paradox.

The main result in this paper shows that there is a competing effect which stops us from taking this convenient limit.
In order for field theory modes from the above region to propagate out and become asymptotic quanta, they must go through geometric uncertainties, like the double-shells in our example, at the distance scale $l$ from the horizon.
If we again fix $l$ and increase $M$, our result in Eq.~(\ref{fin-result}) shows that the unitary propagation to asymptotic Hawking quanta becomes less and less trustworthy. 
Since the dimensionless ratio $(l/M)$ is the only relevant physical parameter that comes into the scaling both effects, there seems to be no middle-ground that can make both unitarity and local-Minkowski-ness arbitrarily reliable.
We think such competition is a good sign and will take us closer to the solution of information paradox.

Let us assume that indeed the QFT-subsystem unitarity is lost during the propagation of Hawking quanta.
This alone does not resolve all the problems. 
Usually, one pictures a non-unitarity process as losing information. 
In the context of black hole information, a Hawking quantum actually has to {\it recover} information. 
A near-horizon mode starts from a maximally mixed (thermal) state with no information, since its partner is the interior mode that never comes out of the horizon. 
After the non-unitarity propagation to become an asymptotic Hawking quantum, it has to carry information to purify the rest of the Hawking radiation there.

Since unitarity is lost to the local geometric uncertainty, what restores information must be the interaction with the local geometry. 
An abstract realization of this information-recovery process is recently demonstrated in \cite{OsuPag16}. 
In order for local geometry to give this information to the Hawking quanta, it must first carry such information.
It was first suggested that superpositions of different classical geometries can store such information \cite{NomVar12}.
In the conventional picture, classical geometries are determined by only a few parameters such as the location and size of the black hole.
Thus the major objection to such picture is that the superpositions of classical geometries do not form a large enough Hilbert space to hold the required information.
Recently, it was realized that there are actually a much larger number of classical geometries that differ by their soft hairs \cite{HawPer16}.
A simple counting suggested that the size of Hilbert space seems no longer the problem, so one should pursue this line of thoughts even further.

Classically, soft hairs manifest themselves as memory effects that alter the distances between geodesics, which is very similar to the effect of geometric uncertainty as we showed in this paper.
Two neighbouring null rays simply ``remember'' the geometric uncertainty they passed through.
Na\"ively speaking, a propagating mode of a massless field should remember a similar effect. 
Based on this picture, one may try to quantize the soft hairs and build a toy model that involves quantum interaction, therefore entanglement, between the soft hairs and the usual QFT modes.
That would provide a complete picture that the loss of subsystem unitarity is really just entanglement with another subsystem which was originally (and illegitimately, as we argued) assumed as the classical background.

\acknowledgments

We thank Raphael Bousso, Pei-Ming Ho and Robert Jefferson for discussions. 
This work is supported by the Canadian Government through the Canadian Institute for Advance Research and Industry Canada, 
and by Province of Ontario through the Ministry of Research and Innovation. 
The Dunlap Institute is funded through an endowment established by the David Dunlap family and the University of Toronto.

\appendix

\section{Calculation of time shift at $r_0\gg M$}
\label{Dtime}

The background geometry is given by Eq.~(\ref{eq-metric}).
An out-going radial null ray is given by
\begin{equation}
	t - t_e = r - r_e + 2M \ln \left( \frac{r - 2M}{r_e - 2M}  \right)~,
	\label{3}
\end{equation}
where $r_e$ and $t_e$ are the starting point of this null ray.
A second null ray which reaches some asymptotic $r_0\gg M$ at a later time is simply
\begin{equation}
	t - (t_e + T) = r - r_e + 2M \ln \left( \frac{r - 2M}{r_e - 2M}  \right)~.
\end{equation}
Clearly, they have the same coordinate time separation $\Delta t = T$ everywhere, as we illustrate in the lower part of Figure \ref{fig-setup}.

Next we introduce two shells of matter with equal and opposite ADM mass, $\Delta M$ and $-\Delta M$ where $|\Delta M| \ll M$, at distances $r_1 < r_2$ as shown in the top part of figure \ref{fig-setup}. 
We model these shells as infinitesimally thin and therefore use the Israel Junction Conditions (IJC) \cite{Isr66} with a delta function shell to analyze the effects of the shells. 
The metric remains the same as Eq.~(\ref{eq-metric}), except for the region between the two shells, which is given by
\begin{equation}
	ds^2_{\rm region \ II} = - \left( 1 - \frac{2M'}{r} \right) d t'^2 + \left( 1 - \frac{2M'}{r} \right)^{-1} dr^2 + r^2 d \Omega_2^2~.	\label{2}
\end{equation}
Here $M' = M + \Delta M$. 
Due to spherical symmetry, we can demand that the angular and radial coordinates to be the same as the background metric. 
But the time coordinate will be different, so we denote that as $t'$.

At the two junctions, the proper time on both sides should be identical, which is the zeroth-order IJC---continuity of spacetime.
That means
\begin{eqnarray}
	\sqrt{1-\frac{2M}{r_1}} \delta t_1 &=& \sqrt{1-\frac{2M'}{r_1}} \delta t'_1~, \\
         \sqrt{1-\frac{2M}{r_2}} \delta t_2 &=& \sqrt{1-\frac{2M'}{r_2}} \delta t'_2~.
\end{eqnarray}

Therefore, the two null rays starting with a separation $T$ at $r_e<r_1$, going through this geometric uncertainty, will end up having a different separation given by
\begin{equation}
T+\Delta T = \sqrt{\frac{r_2-2M}{r_1-2M}} \sqrt{\frac{r_1-2(M+\Delta M)}{r_2-2(M+\Delta M)}}~.
\end{equation}
This clearly leads to Eq.~(\ref{eq-result}):
\begin{equation}
	\frac{\Delta T}{T} 
	= \left( \left( \frac{r_2 - 2M}{r_1 - 2M} \right)^\frac{1}{2} \left( \frac{r_1 - 2(M+\Delta M)}{r_2 - 2(M+\Delta M)} \right)^\frac{1}{2} - 1\right).	\label{10}
\end{equation}

Assuming $r_1$, $r_2$, $M$, $r_1-2M$ are all much larger than $\Delta M$ and $\Delta r=(r_2-r_1)$, we can expand Eq (\ref{eq-result}) to leading order of these two small quantities.
\begin{eqnarray}
	\frac{\Delta T}{T} & = & 
	\left( \left( 1+ \frac{\Delta r}{r_1 - 2M} \right)^\frac{1}{2} \left( 1 + \frac{2\Delta M}{r_1 - 2M} \right)^\frac{1}{2} \left( 1 + \frac{\Delta r + 2 \Delta M}{r_1 - 2M} \right)^{-\frac{1}{2}} - 1 \right)	\nonumber	\\
	& = & - \frac{1}{8} \left( \frac{\Delta r}{r_1 - 2M} \right)^2 + 
	\frac{1}{2} \frac{\Delta r \Delta M}{(r_1 - 2M)^2} - 
	\frac{1}{2} \left( \frac{\Delta M}{r_1 - 2M} \right)^2 - 
	\frac{1}{4} \frac{( \Delta r + 2 \Delta M)^2}{(r_1 - 2M)^2} + 
	\frac{3}{8} \frac{(\Delta r+ 2 \Delta M)^2}{(r_1 - 2M)^2} 	\nonumber	\\
	& = & \frac{\Delta M \Delta r}{(r_1 - 2M)^2} = \frac{\Delta M \Delta r}{r_1^2} g_{tt}^{-2}.
\end{eqnarray}
	
%
%

\section{Local physical quantities}
\label{finalEq}

Here we will translate the coordinate quantities, $\Delta M, r_1$ and $\Delta r$, into the local physical quantities: $\sigma$ as the co-dimension-one energy density of the shell, $l$ as the physical distance from the horizon, and $\Delta l$ as the physical distance between the two shells.
This is given by the Israel Junction Condition.
\begin{equation}
	8 \pi S^a_b = [K^a_b] - [K]h^a_b~.	\label{IJC}
\end{equation}	
Here $K_{ab}$ is the extrinsic curvature corresponding to the induced metric.
$K = K_{ab} h^{ab}$ is the trace of the extrinsic curvature.
The middle brackets ``[ ]'' means we are calculating the extrinsic curvature from both sides of the junction and take their difference.

The induced metrics for the junction at $r_1$, expressed in the coordinate of region $I$ in Fig.\ref{fig-setup}, is the following.
\begin{equation}
	ds_{\rm induced, \ region \ I}^2 = h_{ab} dy^a dy^b \equiv - \left( 1 - \frac{2M}{r} \right) dt^2 + r^2 d \Omega_2^2.	\label{13}
\end{equation}

Analogously we have the same induced metric written in the coordinate of region $II$,
\begin{equation}
	ds_{\rm induced, \ region \ II}^2 = h'_{ab} dy'^a dy'^b \equiv - \left( 1 - \frac{2M'}{r} \right) dt'^2 + r^2 d \Omega_2^2.	\label{14}
\end{equation}

Matching the induced metrics at a fixed radius $r_1$ imposes the following condition. 

\begin{equation}
	dt'^2 = \left( \frac{1 - \frac{2M}{r_1}}{ 1- \frac{2M'}{r_1}} \right) dt^2~,
\end{equation}

We can explicitly calculate the extrinsic curvature components.

\begin{equation}
	K_{ab} = \frac{1}{2} (\mathcal{L}_n g_{\alpha \beta}) e^\alpha_a e^\beta_b, \hspace{5mm} e^\alpha_a \equiv \frac{\partial x^\alpha}{\partial y^a},
\end{equation}
where $a,b \in \{t,\theta, \phi \}$ and $\mathcal{L}_n$ is the Lie derivative w.r.t to the normal vector $n_\alpha = g_{rr}^{\frac{1}{2}} \partial_\alpha r = (0, g_{rr}^{\frac{1}{2}}, 0 , 0)$. Lets look at the Lie derivative first

\begin{eqnarray}
	\mathcal{L}_n g_{\alpha \beta} & = & n^\mu \partial_\mu g_{\alpha \beta} + \partial_\alpha n^\mu g_{\mu \beta} + \partial_\beta n^\mu g_{\alpha \mu}	\nonumber	\\
	& = & n^r \partial_r g_{\alpha \beta} + \partial_\alpha n^r g_{r \beta} + \partial_\beta n^r g_{\alpha r}	
\end{eqnarray}

Now the components of the extrinsic curvature are

\begin{equation}
	K_{ab} = \frac{1}{2} (n^r \partial_r g_{\alpha \beta} + \partial_\alpha n^r g_{r \beta} + \partial_\beta n^r g_{\alpha r}) e^\alpha_a e^\beta_b	\label{Ecuv}
\end{equation}

Notice that the since the only variable in the above equation is $r$ only the $\partial_r$ terms will contribute. Since the $a,b$ indices do not run over $r$ only the first term in the expression above will remain. Using this we can calculate the components of the extrinsic curvature. 

\begin{eqnarray}
	K_{tt} & = & \frac{1}{2} (n^r \partial_r g_{tt}) 	\nonumber	\\
	& = & \frac{1}{2} \left( 1 - \frac{2M}{r} \right)^\frac{1}{2}  \left( \frac{2M}{r^2} \right)
\end{eqnarray}
and we know $(K^t_t)^{I} = g^{tt}K_{tt} = \frac{M}{r^2} \left( 1 - \frac{2M}{r} \right)^{-\frac{1}{2}}$, the $I$ index represents a quantity evaluated in the metric $I$ and similarly for $II$. The other components are 

\begin{equation}
	(K^\theta_\theta)^{I} = \frac{1}{r_1} \left( 1 - \frac{2M}{r_1} \right)^\frac{1}{2} = (K^\phi_\phi)^{I}.
\end{equation}
For the extrinsic curvatures in region II, we simply replace $M$ by $M'$.
\\
\\
The time component $S^t_t$ is the surface energy density and can be calculated using Eq (\ref{IJC}).

\begin{eqnarray}
	\sigma = S^t_t & = & \frac{1}{8 \pi} ((K^\theta_\theta)^{I} + (K^\phi_{\phi})^{I} - (K^\phi_\phi)^{II} - (K^\theta_\theta)^{II})	\nonumber	\\
	& = & \frac{1}{4 \pi r_1} \left( \left( 1 - \frac{2M}{r_1} \right)^\frac{1}{2} - \left( 1 - \frac{2M'}{r_1} \right)^\frac{1}{2} \right).
\end{eqnarray}

When $2\Delta M \ll (r_1-2M)$, this is approximately given by 
\begin{eqnarray}
	\sigma  =  -\frac{1}{4 \pi r_1} \left[ \partial_X \left( 1 - \frac{2X}{r_1} \right)^\frac{1}{2} \right]_{X = M} \Delta M
	 =  \frac{\Delta M}{4 \pi r_1^2} g_{tt}^{-\frac{1}{2}}	\label{sig-Dm}
\end{eqnarray}
Next we need to replace the coordinate distance $\Delta r$ with the proper distance $\Delta l$ and the relation between them is given by
\begin{equation}
	\Delta l = \left( 1 - \frac{2M}{r_1} \right)^{-\frac{1}{2}} \Delta r = g_{tt}^{-\frac{1}{2}} \Delta r.	\label{Dl-Dr}
\end{equation}
These equations allow us to replacing $\Delta M$ by $\sigma$ and $\Delta r$ by $\Delta l$ in Eq.~(\ref{eq-result}) to get Eq.~(\ref{eq-inv}). Finally, Eq.~(\ref{Dl-Dr}) can be integrated to give the proper distance $l$ between the inner shell and the horizon.
\begin{eqnarray}
	l & = & \int^{r_1}_{2M} \left( 1 - \frac{2M}{r} \right)^{-\frac{1}{2}} dr	\nonumber	\\
	& = & \sqrt{(r_1 - 2M)r_1} + M \log \left( \frac{ r_1 - M + \sqrt{(r_1 - 2M)r_1}}{M} \right)	~.
	\label{l-r}
\end{eqnarray}
We will be interested in two extreme cases: $M \ll l$ and $M \gg l$. The value of $l$ in these cases are given by
\begin{equation}
l \approx \begin{cases}
 r_1 - 2M & \ \ \ \ \ (M \ll l)~,	\\
 \sqrt{8M(r_1 - 2M)} & \ \ \ \ \ (M \gg l)~.
\end{cases}	
\end{equation}
This allows us to evaluate Eq.~(\ref{eq-inv}) and get Eq.~(\ref{fin-result})

\bibliography{all_active}

\end{document}